\documentclass[
aps,prd,
12pt,
nopreprintnumbers,
showpacs,
eqsecnum,
nofootinbib
]{revtex4-1}

\usepackage{graphicx}
\usepackage{amssymb}

\begin{document}

\title{Isothermal spheres in general relativity and Ho\v{r}ava-type gravity}
\author{Nahomi Kan}\email[]{kan@gifu-nct.ac.jp}
\affiliation{National Institute of Technology, Gifu College,
Motosu-shi, Gifu 501-0495, Japan}
\author{Kiyoshi Shiraishi}\email[]{shiraish@yamaguchi-u.ac.jp}
\affiliation{
Faculty of Science,
Yamaguchi University, Yamaguchi-shi, Yamaguchi 753--8512, Japan}

\begin{abstract}
We construct a toy model for isothermal spheres in
Ho\v{r}ava gravity, which includes Einstein's gravity if a parameter is
appropriately chosen. 
The equations for the isothermal spheres are derived from the partition function
of the gravitating particle system.
We confirm that the Newtonian limit of the system coincides with the
model of the well-known isothermal sphere. 
The stability of the isothermal sphere is found to be sensitive to the energy
density at the center of the sphere.
\end{abstract}

\maketitle

\section{Introduction}
\label{sec1}

An isothermal sphere is an ideal object, which, for a long time, has been
considered as a Newtonian body of a self-gravitating gas as a model
for a star \cite{Chandra}, as well as for a stellar cluster and a galaxy
\cite{BT}.  The isothermal sphere has asymptotic density  
proportional to $r^{-2}$ where $r$ is the distance from the center of a sphere,
if the distance is sufficiently large.%
\footnote{Such a density profile can explain
the flat rotation curves found in many galaxies.}
At the same time, the isothermal sphere is a nice
theoretical laboratory for the self-gravitational system, which has a novel
thermodynamical behavior \cite{Antonov,LW,Padmanabhan,Chavanis2,Chavanis3}.

The dark matter problem in the Universe has been widely discussed, and
several proposals have appeared with candidates from particle physics
\cite{BHS,Profumo}. On the other hand, modified gravity has been proposed as an
alternative to dark matter \cite{CFPS,CL}. Attempts to reproduce the
density profiles of galaxies have been made with both approaches, but at present,
they have not been firmly resolved. Both modified gravity and
unknown matter have the potential to solve the problem.

Modification of the theory of gravity is also required because of its
consistency as a complete quantum theory, and various attempts have been
made, focusing on the behavior of gravity in the UV region. In this paper, we
investigate the properties of the isothermal sphere in Ho\v{r}ava gravity
\cite{Horava1,Horava2}, which is one of the modified gravity theories expected to
be a UV complete theory. 
By considering the isothermal gas sphere, it is easy to
understand whether the relativistic or modified theory of gravity will be crucial
for certain features of the isothermal sphere, which have mainly%
\footnote{Recently, thermodynamics of isothermal spheres in general relativity has
been reported in Ref.~\cite{Chavanis1}.}
been investigated in the Newtonian systems.%
\footnote{Ultrarelativistic isothermal fluid models in the framework of general
relativity and a modified theory have been studied in
Refs.~\cite{SMD,Dadhich1,Dadhich2,DHM}.}   The present study allows us to better
understand, in addition to the structure and nature of self-gravitational systems,
the modified gravity theories and the nature of coupling to matter in a particular
situation.%
\footnote{The studies on stellar structure models in modified gravitational
theories are reviewed in \cite{ORW}.}

One may suspect that the density profile of spheres at large scales can be affected
by the modification, which is only expected at high energy regions.
However, it has  been reported \cite{SVH} that the structure of isothermal spheres
in the ``softened'' gravity is very different from the Newtonian isothermal
spheres. Softened Newton gravity has a small constant scale in the
gravitational potential. Thus, aside from the magnitude of deviations, it is worth
trying to figure out whether some difference appears in the structure of
isothermal spheres in general relativity and modified theories of gravity.
Furthermore, it should be mentioned that analyses of models of celestial bodies
and galaxies according to the fractional law of gravitation have also appeared
recently
\cite{Giusti,GGV,Varieschi1,Varieschi2,Varieschi3,Seidov}, and that the
higher-derivative modification of Newton's law has also been considered more
recently \cite{Lazar}.

The present paper is organized as follows: In Sec.~\ref{sec2} we introduce the
Ho\v{r}ava-type gravity that we consider in this paper; in Sec.~\ref{sec3}
the partition function of relativistic particles coupled with the metric is
discussed; in Sec.~\ref{sec4} we extract the classical equations of motion for
isothermal spheres from the total partition function of the gravitating system;
in Sec.~\ref{sec5} we illustrate the results for the isothermal spheres in
Einstein gravity and in Ho\v{r}ava gravity; Sec.~\ref{sec6} gives our
conclusions.

\section{Ho\v{r}ava-type gravity}
\label{sec2}

Since its proposal by Ho\v{r}ava \cite{Horava1,Horava2}, there have been
various versions of the Ho\v{r}ava gravity model \cite{Wang,NO}.%
\footnote{In addition, there are many modifications and extensions of
Ho\v{r}ava gravity, which are intended to improve the IR instability in the
original case. Please see Ref.~\cite{CMSVZ} and references therein.} In this
paper, we pick out the simplest model among them
\cite{MKSP,Myung,KS}.  We study an ideal isothermal model in this first
step, and thus deal with the simplified Ho\v{r}ava gravity model in this paper.
However, its generalization will be straightforward. The model
can be regarded as a lowest-order corrected higher-derivative theory of gravity 
beyond Einstein gravity.

The Hamiltonian formalism is adopted for the system in the present analysis.
We start with the ADM line element \cite{ADM},
\begin{equation}
ds^2=-N^2dt^2+\tilde{g}_{ij}(dx^i+N^idt)(dx^j+N^jdt)\,,
\label{ADMm}
\end{equation}
where $N$ is the lapse function, $N^i$ ($i, j=1, 2, 3$) is the shift vector, and
$\tilde{g}_{ij}$ is  the spatial metric.
The Hamiltonian is then written in the form
\begin{equation}
H=\int d^3x\,\sqrt{\tilde{g}}\left(N\mathcal{H}+N^i\mathcal{H}_i\right)\,,
\end{equation}
where $\tilde{g}$ is the determinant of the matrix elements $\tilde{g}_{ij}$.
Here, the Hamiltonian constraint $\mathcal{H}$ and the momentum constraint
$\mathcal{H}_i$ are given in the $z=2$ model%
\footnote{The dynamical critical exponent $z$ indicates that the mass dimension of
time is equal to $-z$ \cite{Horava1,Wang}.}
 of Ho\v{r}ava gravity
\cite{Horava1,MKSP},
\begin{equation}
\mathcal{H}=\frac{\kappa^2}{2\sqrt{\tilde{g}}}\left[
\pi^{ij}\pi_{ij}-\frac{\lambda}{3\lambda-1}(\pi^i_i)^2\right]
-\mu^3\tilde{R}-\frac{\kappa^2\mu^2(4\lambda-1)}{32(3\lambda-1)}\tilde{R}^2
+\frac{\kappa^2\mu^2}{8}
\tilde{R}^{ij} \tilde{R}_{ij}\,,
\end{equation}
\begin{equation}
\mathcal{H}_i=-2\nabla^j\pi_{ij}\,,
\end{equation}
with $\pi^{ij}$ being the conjugate momentum of $\tilde{g}_{ij}$, and $\tilde{R}$
and
$\tilde{R}_{ij}$ being the scalar curvature and the Ricci tensor constructed from
$\tilde{g}_{ij}$, respectively. Here, $\nabla_i$ denotes the three-dimensional
covariant derivative, and the parameters
$\kappa$ and
$\lambda$ are constants.%
\footnote{Hereafter, we use the traditional notation as in
\cite{MKSP,Myung,KS}.} In the case of $\lambda=1$, the Hamiltonian density
$\mathcal{H}$ becomes \cite{MKSP,Myung,KS}
\begin{equation}
\mathcal{H}=\frac{\kappa^2}{2\sqrt{\tilde{g}}}\left[
\pi^{ij}\pi_{ij}-\frac{1}{2}(\pi^i_i)^2\right]
-\mu^3\left[\tilde{R}
-\frac{2}{\omega}\left(
\tilde{R}^{ij} \tilde{R}_{ij}-\frac{3}{8}\tilde{R}^2\right)\right]\,,
\end{equation}
with $\omega\equiv\frac{16\mu}{\kappa^2}$. Hereafter, we consider this 
Ho\v{r}ava-type gravity model.

The higher-order terms are important when discussing UV completion of quantum
gravity in general. We focus on the lowest-order deviation
from the general theory of relativity in this model by considering the $z=2$
model.

\section{Grand canonical partition function for relativistic particles in
curved spacetime}
\label{sec3}

We consider the constituent of isothermal spheres as an ideal gas of
noninteracting classical particles (which may be celestial bodies).
One can write the Hamiltonian of the $n$-particle system in the background
spacetime as 
\begin{equation}
H_n=N\sum_{a=1}^n\sqrt{\tilde{g}^{ij}p^a_ip^a_j+m^2}\,,
\end{equation}
where $m$ is the common mass of the particles and $p^a_i$ denotes the momentum of
the $a$th particle located at $q_a$.

Let us consider the grand canonical formalism for the isothermal system.
Then, the grand canonical partition function at temperature $T$ is written as
\cite{GNS}
\begin{eqnarray}
Z_G&=&\sum_{n=0}^\infty\frac{z^n}{n!}\int\int\prod_{a=1}^n\frac{d^3p_ad^3q_a}{(2\pi)^3}
e^{-\beta H_n}\nonumber \\
&=&\sum_{n=0}^\infty\frac{z^n}{n!}\int\prod_{a=1}^n
d^3q_a\,\sqrt{\tilde{g}}\,\frac{m^3}{2\pi^2}\frac{K_2(\beta mN)}{\beta mN}\nonumber
\\ &=&\sum_{n=0}^\infty\frac{z^n}{n!}\left[\int
d^3q\,\sqrt{\tilde{g}}\,\frac{m^3}{2\pi^2}\frac{K_2(\beta mN)}{\beta
mN}\right]^n\nonumber \\ &=&\exp\left[\int
d^3x\,\sqrt{\tilde{g}}\,\frac{z\,m^3}{2\pi^2}\frac{K_2(\beta mN)}{\beta
mN}\right]\,,
\end{eqnarray}
where $\beta=1/T$ and $z$ is the activity. The special function $K_\nu(z)$ is the
modified Bessel function of the second kind.
Here, the background metric is assumed to be fixed or nearly constant, as we
consider only the adiabatic situation or local equilibrium. It should be noted
that, in the limiting cases, the expression reduces to
\begin{equation}
\frac{z\,m^3}{2\pi^2}\frac{K_2(\beta mN)}{\beta
mN}\approx\left\{
\begin{array}{ll}
z\left(\frac{m}{2\pi\beta N}\right)^{3/2}e^{-\beta mN} & \quad\beta mN\gg 1\\
\frac{z}{\pi^2(\beta N)^3} & \quad\beta mN\ll 1\,.
\end{array}
\right.
\end{equation}
It is also noteworthy that the inverse temperature $\beta$ appears
only as the combination $\beta N=\beta\sqrt{-g_{00}}$, as advocated by Tolman
\cite{Tolman}.

Using the partition function, we find  that
the particle number density is given by 
\begin{equation}
n_p=z\frac{\partial}{\partial z}\left[
\frac{z\,m^3}{2\pi^2}\frac{K_2(\beta mN)}{\beta
mN}\right]=\frac{z\,m^3}{2\pi^2}\frac{K_2(\beta mN)}{\beta mN}\,,
\end{equation}
while the pressure of the gas is given by
\begin{equation}
P=\frac{z\,m^3}{2\pi^2}\frac{K_2(\beta mN)}{\beta^2
mN^2}=\frac{n_p}{\beta N}\,,
\label{PP}
\end{equation}
and the energy density is given by
\begin{equation}
\rho=-\frac{1}{N}\frac{\partial}{\partial\beta}\left[
\frac{z\,m^3}{2\pi^2}\frac{K_2(\beta mN)}{\beta
mN}\right]=\frac{z\,m^3}{2\pi^2\beta N}\left[K_1(\beta mN)+ \frac{3
K_2(\beta mN)}{\beta mN}\right]\,.
\label{rhorho}
\end{equation}
It is easy to see that, in the well-known relativistic limit $\beta m\ll 1$,
the equation of state becomes $\rho=3P$.
Note that $P/\rho\approx 0.01$ for $\beta mN=100$,
$P/\rho\approx 0.08$ for $\beta mN=10$, and
$P/\rho\approx 0.1$ for $\beta mN=6$.

\section{Equations for isothermal spheres from the partition function}
\label{sec4}

We consider the total adiabatic system of 
gravity coupled to isothermal gas.  The partition function can be represented by
the path integral of the variables if the  spacetime is approximately static. We
assume that the
shift vector vanishes for nonrotating bodies, and the
integration over conjugate momentum
$\pi^{ij}$ is omitted.%
\footnote{In other words, the graviton degrees of freedom are out of thermal
equilibrium.}
Then, the grand canonical partition function in this system is written as
\begin{equation}
Z_G=\int[DN][D\tilde{g}_{ij}]\exp\left\{\int\left[-\beta N{}\overline{\mathcal{H}}
+\frac{z m^3}{2\pi^2}\frac{K_2(\beta mN)}{\beta mN}
\right]\sqrt{\tilde{g}}d^3x\right\}\,, 
\label{eq20}
\end{equation}
where
\begin{equation}
\overline{\mathcal{H}}=
-\mu^3\left[\tilde{R}
-\frac{2}{\omega}\left(
\tilde{R}^{ij} \tilde{R}_{ij}-\frac{3}{8}\tilde{R}^2\right)\right]\,.
\label{ZZ}
\end{equation}

We can derive equations for the static equilibrium configuration.
Such equations are obtained by the evaluation of the steepest descent or the
variation of the total Hamiltonian, which is described by the exponential in
(\ref{ZZ}). One can obtain the following classical equations of motion from the
variational principle:
\begin{eqnarray}
& &\tilde{R}
-\frac{2}{\omega}\left(
\tilde{R}^{ij} \tilde{R}_{ij}-\frac{3}{8}\tilde{R}^2\right)=\frac{zm}{\mu^3}
\frac{m^3}{2\pi^2}\frac{1}{\beta mN}\left[K_1(\beta
mN)+\frac{3}{\beta mN}K_2(\beta
mN)\right]=\frac{1}{\mu^3}\rho\,,\label{NN}\\
& &N\left\{\tilde{R}_{ij}-\frac{1}{2}\tilde{R}\tilde{g}_{ij}-\frac{2}{\omega}\left[
2\tilde{R}_{ik}\tilde{R}_j^k-\frac{1}{2}\tilde{R}_{kl}\tilde{R}^{kl}\tilde{g}_{ij}
-\frac{3}{8}\left(2\tilde{R}\tilde{R}_{ij}-\frac{1}{2}\tilde{R}^2\tilde{g}_{ij}\right)\right]
\right\}\nonumber \\
& &-\nabla_i\nabla_jN+\nabla^2N\tilde{g}_{ij}-\frac{2}{\omega}\Biggl[
-\nabla^k\nabla_i(N\tilde{R}_{jk})-\nabla^k\nabla_j(N\tilde{R}_{ik})
+\nabla_k\nabla_l(N\tilde{R}^{kl})+\nabla^2(N\tilde{R}_{ij})\nonumber \\
& &-\frac{3}{8}
\left(-2\nabla_i\nabla_j(N\tilde{R})+2\nabla^2(N\tilde{R})\tilde{g}_{ij}\right)
\Biggr]=\frac{z}{2\beta \mu^3}\frac{m^3}{2\pi^2}\frac{K_2(\beta
mN)}{\beta mN}\tilde{g}_{ij}=\frac{N}{2\mu^3}P\tilde{g}_{ij}\,,\label{EE}
\end{eqnarray}
where $\nabla^2\equiv\nabla_k\nabla^k$.
Incidentally, the trace of (\ref{EE}) gives
\begin{eqnarray}
& &-\frac{1}{2}N\tilde{R}+2\nabla^2N-\frac{2}{\omega}\left[\frac{1}{2}N
\left(\tilde{R}^{ij} \tilde{R}_{ij}-\frac{3}{8}\tilde{R}^2\right)+
(\nabla_k\nabla_lN)\left(\tilde{R}^{kl}-\frac{1}{2}\tilde{R}\tilde{g}^{kl}\right)
\right]\nonumber \\
&=&\frac{3z}{2\beta \mu^3}\frac{m^3}{2\pi^2}\frac{K_2(\beta
mN)}{\beta mN}=\frac{3N}{2\mu^3}P\,.\label{TT}
\end{eqnarray}
Note that the general relativistic case can be obtained if
$\frac{2}{\omega}\rightarrow 0$. Then, Eqs.~(\ref{NN}) and (\ref{TT}) give
the formal ``classical'' equation for $N$: 
\begin{equation}
\frac{1}{N}\nabla^2N=\frac{z}{4\mu^3}\frac{m^4}{2\pi^2(\beta
mN)}\left[K_1(\beta mN)+\frac{6}{\beta mN}K_2(\beta mN)\right]
=\frac{1}{4\mu^3}(\rho+3P)\,.\label{GP}
\end{equation}
The Newtonian limit is attained if $N^2\approx 1+2\phi$, $\tilde{g}_{ij}\approx
\delta_{ij}$, $\beta m\gg 1$, and $\mu^3=\frac{1}{16\pi G}$, where $G$ is 
Newton's constant. Keeping the lowest order terms, Eq.~(\ref{GP}) leads to
\begin{equation}
\nabla^2\phi=4\pi G\rho_0 e^{-\beta
m\phi}\,,
\end{equation}
where we set $\rho_0\equiv ze^{-\beta m}m\left(\frac{m}{2\pi\beta}\right)^{3/2}$.
This equation is already known for the Newtonian isothermal gas
\cite{Chandra,BT,Antonov,LW,Padmanabhan,Chavanis2,Chavanis3}.

Now, we discuss the case with spherical symmetry.
If we assume static, spherically symmetric space, we can take $N_i=0$ in the ADM
line element (\ref{ADMm}). Then, the metric becomes
\begin{equation}
ds^2=-N^2(r)\,dt^2+\frac{dr^2}{1-\frac{2GM(r)}{r}}+r^2(d\theta^2+\sin^2\theta\,
d\varphi^2)\,,
\label{metric}
\end{equation}
where $G=\frac{1}{16\pi\mu^3}$ is Newton's constant. The function $M(r)$ 
describes the mass inside the sphere with radius $r$.%
\footnote{Since the equation from the variation of $N$ gives $M'$, $M$ is
proportional to the volume integral of the $00$ component of the energy-momentum
tensor, which appears in the right-hand side of the Einstein equation.}
Substituting the metric (\ref{metric}), the equations of motion are reduced to
\begin{eqnarray}
& &\frac{G}{r^2}\left[1+\frac{2}{\omega}\frac{GM(r)}{r^3}\right]\frac{dM(r)}{dr}
-\frac{2}{\omega}\frac{3G^2M^2(r)}{2r^6}
=4\pi G\rho(r)\,, \\ &
&\frac{1}{r}\left[1-\frac{2GM(r)}{r}\right]\left[1+\frac{2}{\omega}\frac{GM(r)}{r^3}\right]\frac{dN(r)}{dr}
-\frac{G}{r^3}\left[1-\frac{2}{\omega}\frac{GM(r)}{2r^3}\right]M(r)N(r)
\nonumber \\
& &=4\pi G N P(r)\,,
\end{eqnarray}
where $\rho(r)$ and $P(r)$ are defined by (\ref{rhorho}) and (\ref{PP}) with
$N\rightarrow N(r)$. Note that the second or higher derivatives of the functions
are eliminated. In order to simplify the equations further, we rescale the
variables,
\begin{equation}
x\equiv\sqrt{4\pi G \rho_c}\, r\,,\quad
y\equiv\beta mN\,,\quad
\tilde{M}\equiv\sqrt{4\pi G \rho_c}GM\,,\quad
\alpha\equiv 4\pi G \rho_c\frac{2}{\omega}\,,
\end{equation}
with
\begin{equation}
\rho_c\equiv\frac{z\,m^4}{2\pi^2 y_0}\left[K_1(y_0)+ \frac{3
K_2(y_0)}{y_0}\right]\,,\quad y_0\equiv y(0).
\label{rho}
\end{equation}
Then, the equations for $y(x)$ and $\tilde{M}(x)$ become
\begin{eqnarray}
& &\frac{1}{x^2}\left(1+\alpha\frac{\tilde{M}}{x^3}\right)\tilde{M}'
-\alpha\frac{3\tilde{M}^2}{2x^6}=\frac{y_0}{y}
\frac{K_1(y)+ \frac{3
K_2(y)}{y}}{K_1(y_0)+ \frac{3
K_2(y_0)}{y_0}}\,,\label{eex1}
\\ &
&\frac{1}{x}\left(1-\frac{2\tilde{M}}{x}\right)\left(1+\alpha\frac{\tilde{M}}{x^3}
\right)y'
-\frac{1}{x^3}\left(1-\alpha\frac{\tilde{M}}{2x^3}\right)\tilde{M}\,y
=\frac{y_0}{y}
\frac{K_2(y)}{K_1(y_0)+ \frac{3
K_2(y_0)}{y_0}}\,,
\label{eex2}
\end{eqnarray}
where the prime (${}'$) means the derivative with respect to $x$.
We must find solutions satisfying the boundary conditions
\begin{equation}
\tilde{M}(0)=0\,,\quad y(0)=y_0\,.
\end{equation}

In the next section, we exhibit the numerical results.

\section{Numerical calculations}
\label{sec5}

\subsection{Isothermal spheres in Einstein gravity}

First, we consider isothermal spheres in Einstein gravity, i.e., in the case with
$\alpha=0$.
We define two functions:
\begin{equation}
u\equiv \frac{d\ln M(r)}{d\ln
r}=x\frac{\tilde{M}'(x)}{\tilde{M}(x)}\,,
\quad
v\equiv \beta m\frac{dN(r)}{d\ln r}=x y'(x)\,.
\end{equation}
Note that the Newtonian limit of $v$ yields
\begin{equation}
v=\beta m r\frac{dN(r)}{dr}=(\beta m N) \frac{r}{N(r)}\frac{dN(r)}{dr}
\rightarrow
\frac{\rho(r)}{P(r)}r\frac{d\phi(r)}{dr}\approx\frac{\rho(r)}{P(r)}\frac{GM(r)}{r}\,,
\end{equation}
where $\phi$ denotes the Newtonian gravitational potential. 

\begin{figure}[ht]
\centering
\includegraphics
{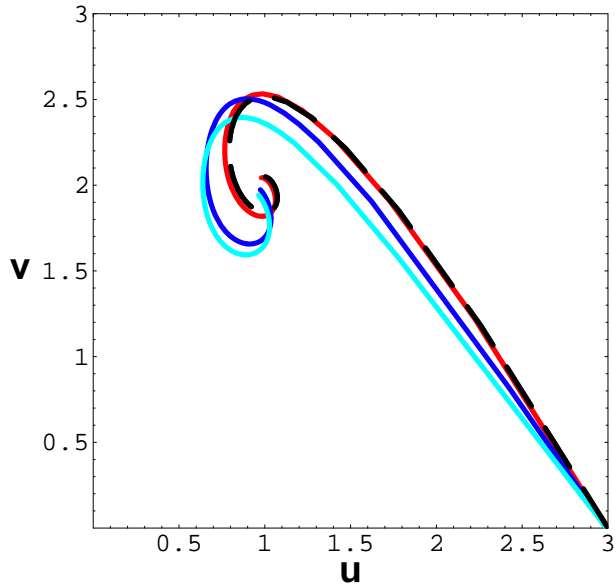}
\caption{The $(u, v)$ curves for $y_0=100$ (red), $y_0=10$ (blue), and $y_0=6$
(cyan). The black dashed curve indicates the Newtonian isothermal sphere.}
\label{fig1}
\end{figure}

In Fig.~\ref{fig1}, we show the solutions for various initial conditions in the
$(u, v)$ plane.
The black dashed curve indicates the Newtonian isothermal sphere \cite{Chandra}.
The curves in red, blue, and cyan correspond to the boundary conditions $y_0=100$,
$10$, and $6$, respectively.
All the curves start at the point $(u, v)=(3,0)$, which represents the center of
the isothermal sphere, and they approach the fixed point $(u, v)\approx (1,2)$,
which corresponds to $x\rightarrow \infty$.
In our present model, since the equation of state in the asymptotic region $x\gg 1$
becomes nonrelativistic as the density decreases,
the behavior $M(r)\propto r$ is the same as in the case of the Newtonian
isothermal sphere \cite{Chandra,BT,Antonov,LW,Padmanabhan,Chavanis2,Chavanis3}.

\begin{figure}[ht]
\centering
\includegraphics
{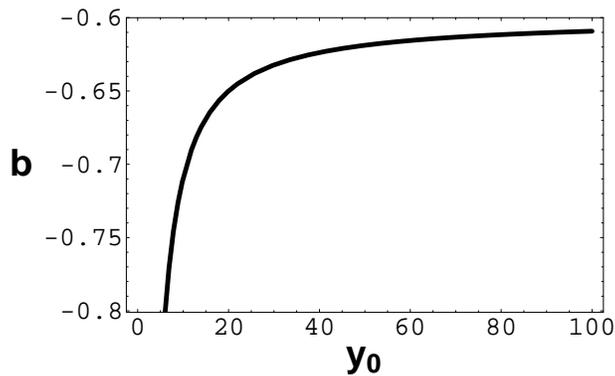}
\caption{Slope coefficient $b$ versus $y_0$.}
\label{fig2}
\end{figure}

The behavior of curves near $(u, v)=(3,0)$ is found to be
\begin{equation}
u=3+b v\,\quad (v\ll 1)\,,
\end{equation}
where
\begin{equation}
b=-\frac{3\left[\frac{3}{y_0}K_1(y_0)+\left(1+\frac{12}{y_0^2}\right)K_2(y_0)\right]}{5
\left[K_1(y_0)+\frac{3}{y_0}K_2(y_0)\right]}
\approx\left\{\begin{array}{ll}
-\frac{3}{5} & (y_0\gg 1)\\
-\frac{12}{5 y_0} & (y_0\ll 1)
\end{array}
\right.\,.
\end{equation}
Figure \ref{fig2} shows $b$ plotted against $y_0$.

\begin{figure}[ht]
\centering
\includegraphics
{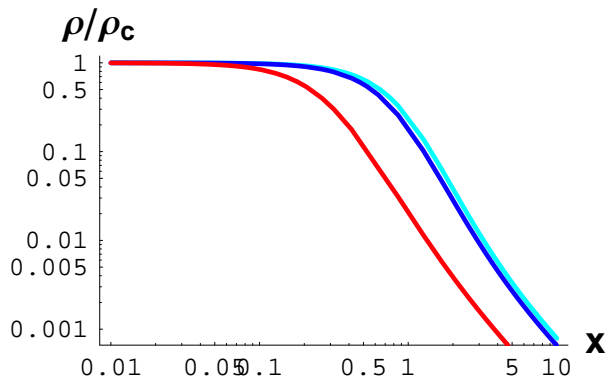}
\caption{Density profiles $\rho(x)/\rho_c$ of general relativistic
isothermal spheres for
$y_0=100$ (red),
$y_0=10$ (blue), and $y_0=6$ (cyan). }
\label{fig3}
\end{figure}

Figure \ref{fig3} shows the density profiles $\rho(x)/\rho_c$ as functions of
$x\propto r$, where the curves in red, blue, and cyan correspond to the boundary
conditions
$y_0=100$,
$10$, and $6$, respectively.
In all of these cases, we find the asymptotic behavior $\rho\propto 1/r^2$,
similarly to that of the Newtonian isothermal sphere which extends to infinity.
Incidentally, it turns out that the asymptotics $v\approx 2$ read $y\approx 2\ln
x$.

\subsection{Isothermal spheres in Ho\v{r}ava gravity}

Next, we consider the isothermal spheres in Ho\v{r}ava gravity.

\begin{figure}[ht]
\centering
\includegraphics[width=5cm]{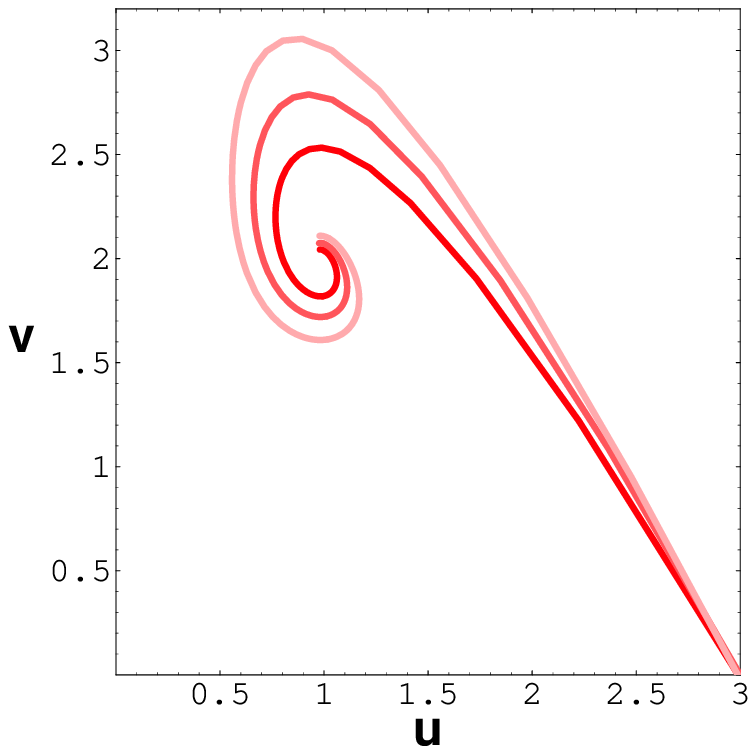}\quad
\includegraphics[width=5cm]{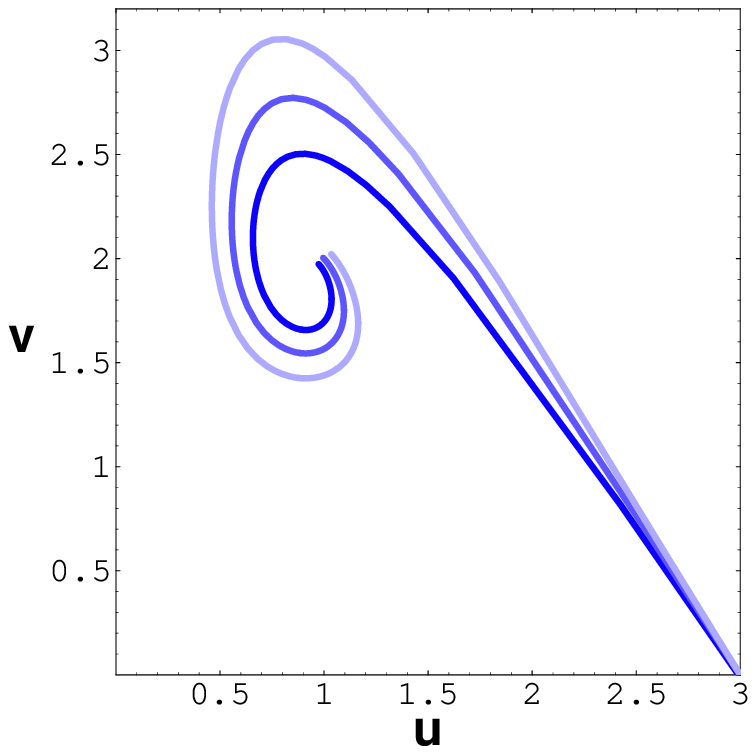}\quad
\includegraphics[width=5cm]{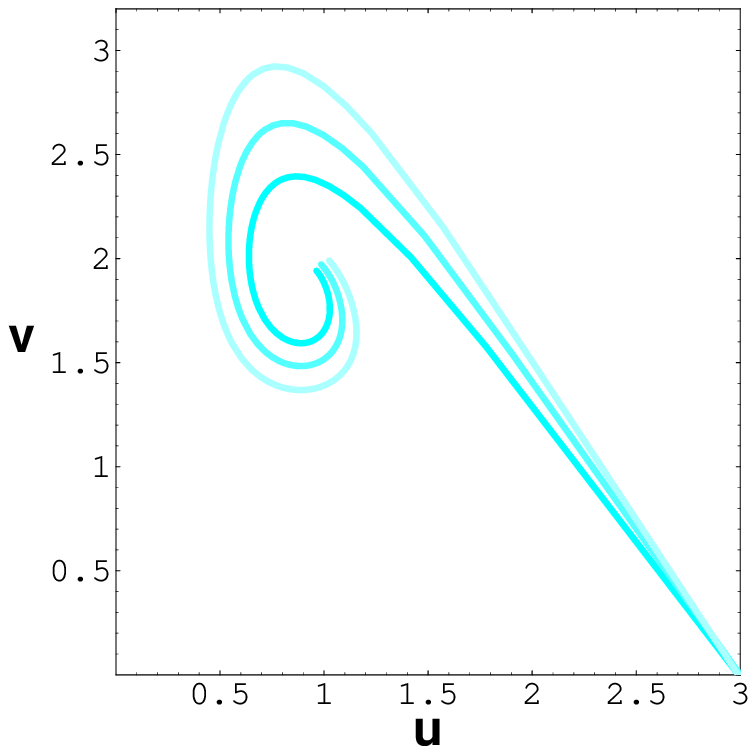}\\
(a) \hspace{5cm} (b) \hspace{5cm} (c)
\caption{Plots of spirals in the $(u,v)$ plane. (a) $y_0=100$, (b) $y_0=10$, and
(c) $y_0=6$. From the inner spiral to the outer spiral, $\alpha=0$,
$\alpha=0.5$, and $\alpha=1$ in each plot.}
\label{fig4}
\end{figure}

The spirals of solutions in the $(u,v)$ plane are shown in Fig.~\ref{fig4}.
The fixed point $(u,v)\approx (1,2)$ is almost unchanged.

Near the starting point, $(u,v)=(3,0)$, the curve is approximated as
\begin{equation}
u=3+b v\,\quad (v\ll 1)\,,
\end{equation}
where
\begin{equation}
b=-\frac{3
\left(1+\frac{2\alpha}{3}+\sqrt{1+\frac{2\alpha}{3}}\right)}{10
\left(1+\frac{2\alpha}{3}\right)}\frac{\frac{3}{y_0}K_1(y_0)+
\left(1+\frac{12}{y_0^2}\right)K_2(y_0)}{
K_1(y_0)+\frac{3}{y_0}K_2(y_0)}
\approx\left\{\begin{array}{ll}
-\frac{3
\left(1+\frac{2\alpha}{3}+\sqrt{1+\frac{2\alpha}{3}}\right)}{10
\left(1+\frac{2\alpha}{3}\right)} & (y_0\gg 1)\\
-\frac{6
\left(1+\frac{2\alpha}{3}+\sqrt{1+\frac{2\alpha}{3}}\right)}{5
\left(1+\frac{2\alpha}{3}\right)y_0} & (y_0\ll 1)\,.
\end{array}
\right.
\end{equation}

\begin{figure}[ht]
\centering
\includegraphics
{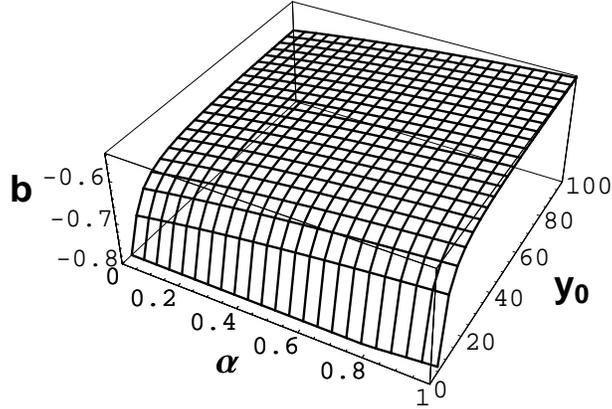}
\caption{Slope coefficient $b$ plotted against $\alpha$ and $y_0$.}
\label{fig5}
\end{figure}

Thus, the spirals become larger according to the increase in the value of $\alpha$.
The curves in the low-density case with $y_0=100$ are given in Fig.~\ref{fig4}(a).
The relatively high-density cases with $y_0=10$ [Fig.~\ref{fig4}(b)]
and $y_0=6$ [Fig.~\ref{fig4}(c)] exhibit similar characteristics.
The coefficient  $b$ is plotted against $\alpha$ and $y_0$ in Fig.~\ref{fig5}.

\begin{figure}[ht]
\centering
\includegraphics[width=7cm]{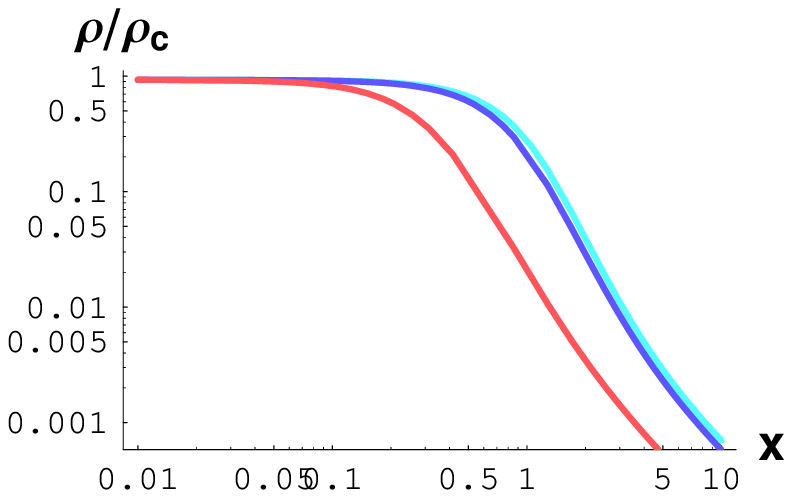}\qquad
\includegraphics[width=7cm]{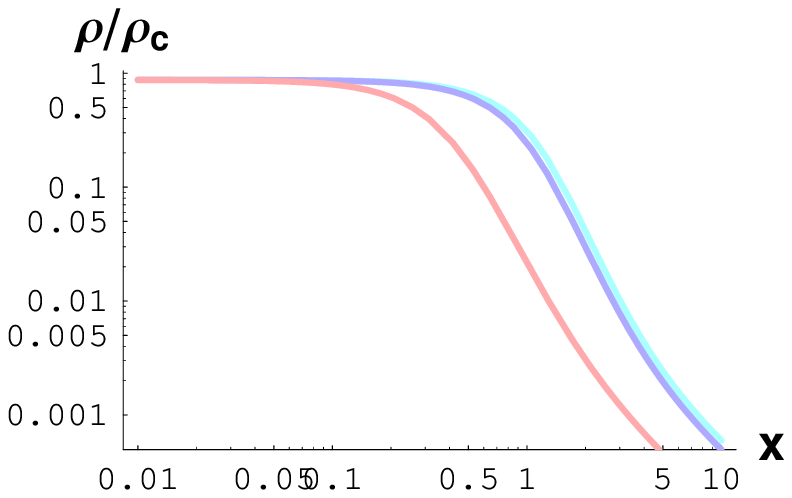}\\
(a) \hspace{7cm} (b)
\caption{The density profiles $\rho(x)/\rho_c$ for $y_0=100$ (red), $y_0=10$
(blue), and $y_0=6$ (cyan) with the parameter $\alpha=0.5$ are plotted in (a)
while those with $\alpha=1$ are plotted in (b).}
\label{fig6}
\end{figure}

The profiles of the energy density for $\alpha=0.5$ and $1$ are shown in
Fig.~\ref{fig6}. Note that here $\rho$ is defined by $\propto \frac{1}{4\pi r^2
}\frac{dM(r)}{dr}$. They seem to have behaviors similar to the case in general
relativity ($\alpha=0$). This is because the terms with the coefficient
$\alpha$ in the equations are proportional to $M/r^3$, which behaves
asymptotically
$\propto 1/r^2$ for $r\rightarrow \infty$.
In other words, all isothermal spheres in our present model have an outer region
which is well described by the structure of the Newtonian isothermal spheres.

\subsection{Stability}

While the behaviors of the spirals in the $(u,v)$ plane
and the density profiles have moderate dependence on the central density and the
parameter $\alpha$ which appears in Ho\v{r}ava gravity,
the parameter dependence of stability is very complicated as we will show below.
Therefore, in the present paper, we only discuss the stability by considering the
ratio of the sum of the mass of constituent particles and the mass of the
isothermal sphere in the region of the fixed radius. The analyses using various
known methods are left for future studies.

Because the isothermal spheres in our model have the same asymptotic density
profile as the Newtonian one, we consider the finite spherical box to define the
mass of the object \cite{Chandra,BT,Antonov,LW,Padmanabhan,Chavanis2,Chavanis3}.
We consider the region inside the sphere with radius $r$.

Here, we consider the ratio $mN_p/M$, where
\begin{equation}
N_p=4\pi\int_0^r n_p(r) \frac{r^2}{\sqrt{1-\frac{2GM(r)}{r}}}
dr=\frac{1}{G\sqrt{4\pi G\rho_c}}\int_0^x
\frac{y_0
K_2(y(x'))}{y(x')[(K_1(y_0)+\frac{3}{y_0}K_2(y_0)]}\frac{{x'}^2}{\sqrt{1-
\frac{2\tilde{M}(x')}{x'}}}dx'\,.
\end{equation}
Then, the ratio can be expressed as
\begin{equation}
\frac{m N_p}{M}=\frac{1}{\tilde{M}(x)}\int_0^x
\frac{y_0 K_2(y(x'))}{y(x')[(K_1(y_0)+\frac{3}{y_0}K_2(y_0)]}\frac{{x'}^2}{\sqrt{1-
\frac{2\tilde{M}(x')}{x'}}}dx'\,.
\end{equation}

\begin{figure}[ht]
\centering
\includegraphics[width=5cm]{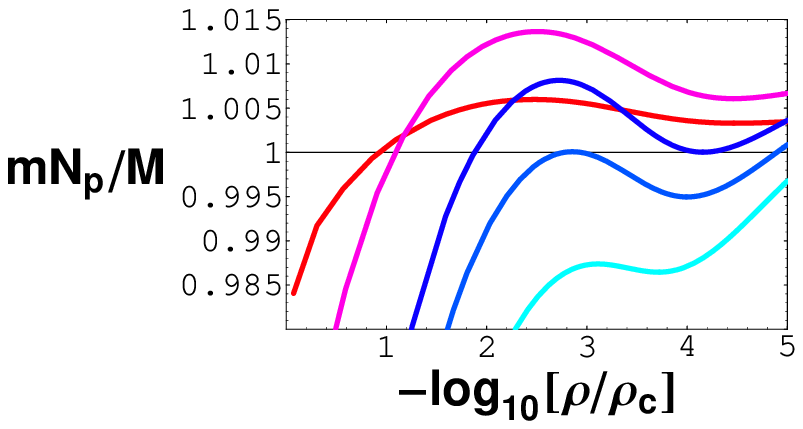}\quad
\includegraphics[width=5cm]{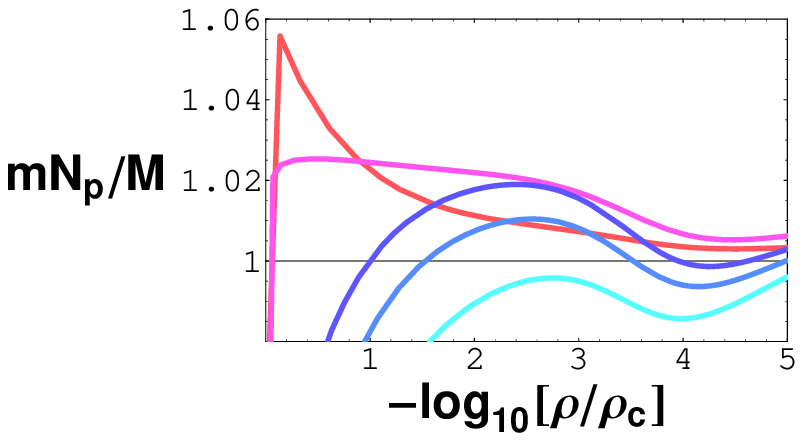}\quad
\includegraphics[width=5cm]{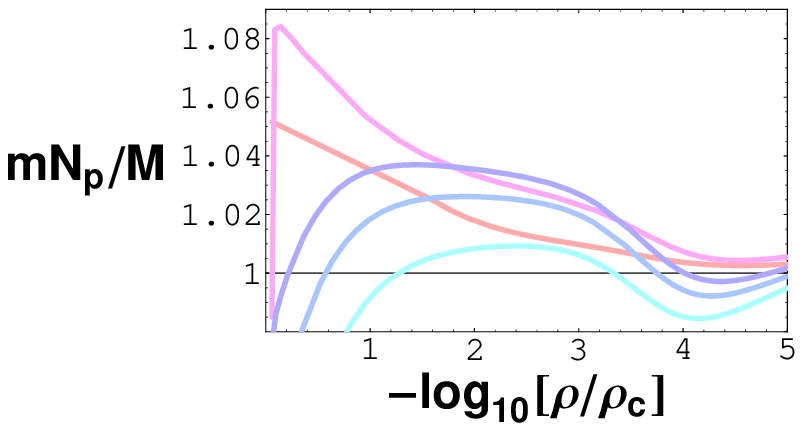}\\
(a) \hspace{5cm} (b) \hspace{5cm} (c)
\caption{The ratio $mN_p/M$ is plotted. (a) $\alpha=0$, (b) $\alpha=0.5$, and
(c) $\alpha=1$. In each plot, the red curve represents $y_0=100$,
the magenta curve is for $y_0=30$,
the blue curve is for $y_0=10$,
the pale blue curve is for $y_0=7.8$, and
the cyan curve is for $y_0=6$.}
\label{fig7}
\end{figure}

Figure \ref{fig7} shows the ratio $mN_p/M$ versus $-\log_{10}[\rho/\rho_c]$.
Since the ratio $\rho(r)/\rho_c$ monotonically decreases with $r$, as we have seen
in this section, the radius of the spherical box becomes larger from left to right
in the horizontal axis. Notice that, apart from some exceptions, the
characteristic feature of the plots lies inside the finite region by selecting the
scale of the axis. 

Since
$N_p=\int n_p
\sqrt{\tilde{g}}d^3x$, the ratio
$mN_p/M>1$ implies positive binding energy $m N_p-M$, so the isothermal sphere
is expected to be energetically stable in this case.

We should also consider the maximum point of the ratio. On the right-hand side of
the maximum, the increase of the radius of the sphere reduces the amount of 
binding energy per mass. Thus, there is a possibility that some smaller
bodies produced by fission will be more stable than a single body.

In Fig.~\ref{fig7}(a), we show the general relativistic case $(\alpha=0)$.
If the central density is sufficiently large (i.e., the gas is rather relativistic
in the vicinity of the center), the stable configuration disappears according to
the above-mentioned criteria. Its critical value is $y_0\approx 7.8$.
However, the change of the curve is complicated for $y_0> 10$.
For $y_0=100$, the maximum is located around $-\log_{10}[\rho(r)/\rho_c]\approx
2.5$. This is consistent with the known stability criterion for the Newtonian
isothermal sphere $-\log_{10}[\rho(r)/\rho_c]<2.85$
\cite{Antonov,LW,Padmanabhan,Chavanis2,Chavanis3}.

The Ho\v{r}ava-type gravity case is more complicated.
The behaviors of the curves in Figs.~\ref{fig7}(b) (for $\alpha=0.5$) and
\ref{fig7}(c) (for
$\alpha=1$) drastically change if $y_0$ is larger than about $10$.
This is because the terms including $\alpha$ in (\ref{eex1}) and (\ref{eex2}),
which have finite values in the central region [i.e., $\tilde{M}(x)/x^3=
\frac{2}{3+\sqrt{9+6\alpha}}+O(x^2)$ if $x\ll 1$], are dominant over the
contribution of densities in the right-hand side of the equations in the small
scale. Thus, the almost nonrelativistic gas sphere is stable when the radius is
relatively small. 
Another important feature one can observe in Fig.~\ref{fig7} is that 
even the relatively high-central-density spheres may possess a stable radius if
$\alpha$ is sufficiently large.

\section{Conclusion}
\label{sec6}

In this paper, we investigated isothermal spheres in general relativity and in a
concise version of the Ho\v{r}ava gravity model. 
We concentrated on spherically symmetric and static solutions of the
equations for the local equilibrium configuration. 
We found that the nonrelativistic limit reproduces the already-known Newtonian
isothermal sphere.

We found that with Einstein gravity
the stability of the sphere tends to be spoiled by the high density
at the center of the general relativistic isothermal sphere.
In Ho\v{r}ava-type gravity, the higher-derivative term stabilizes the sphere 
even if the central density is rather high; however, at the same time, the term
makes the radius of the stable sphere very small with low central density.

In vacuum, the value of the parameter of Ho\v{r}ava gravity is
severely limited observationally \cite{HKL}. 
However, recalling that the parameter
$\alpha$ is defined as $4\pi G\rho_c\frac{2}{\omega}$,
it can be found that the value of $\alpha$ is generally enhanced in
matter. Thus, it is meaningful to consider the higher-derivative correction 
in the study of the general high-density stellar structure.
Studying various modified gravitational theories and other choices of
equations of state is a good direction for future work.
We will also continue discussing the stability from various points of view,
including thermodynamical aspects, and we shall report such analyses in a future
work.

Last but not least, we approached finite-temperature
self-gravitating system from first principles.  
As a future task, the analyses of thermodynamical fluctuations with metric
fluctuations  should be studied carefully for investigating thermodynamical
quantities in the system. A further 
theoretical study, including an extension to nonextensive
statistical dynamics \cite{TS1,TS2,TS3,Chavanis4,Chavanis5} and generalization of
canonically formulated gravity \cite{MN,AFMNOP,YOGM},
 will be reported elsewhere.


\acknowledgments
We would like to thank P.~H.~Chavanis and N.~Dadhich for providing information
about their work on isothermal spheres, and S.~Vagnozzi for providing information
on modified Ho\v{r}ava gravity.

\bibliographystyle{apsrev4-1}


\end{document}